\documentclass[aip,amsmath,amssymb,reprint,twocolumn]{revtex4-1}

\usepackage{graphicx,hyperref}

\begin{document}

% Use the \preprint command to place your local institutional report number 
% on the title page in preprint mode.
% Multiple \preprint commands are allowed.
%\preprint{}

\title{The Infinite Interface Limit of Multiple-Region Relaxed MHD} %Title of paper

% repeat the \author .. \affiliation  etc. as needed
% \email, \thanks, \homepage, \altaffiliation all apply to the current author.
% Explanatory text should go in the []'s, 
% actual e-mail address or url should go in the {}'s for \email and \homepage.
% Please use the appropriate macro for the type of information

% \affiliation command applies to all authors since the last \affiliation command. 
% The \affiliation command should follow the other information.

\author{G.R. Dennis}
\email[]{graham.dennis@anu.edu.au}
%\homepage[]{Your web page}
%\thanks{}
%\altaffiliation{}
\affiliation{Research School of Physics and Engineering, Australian National University, ACT 0200, Australia}

\author{S.R. Hudson}
\affiliation{Princeton Plasma Physics Laboratory, PO Box 451, Princeton, NJ 08543, USA}

\author{R.L. Dewar}
\author{M.J. Hole}
\affiliation{Research School of Physics and Engineering, Australian National University, ACT 0200, Australia}

% Collaboration name, if desired (requires use of superscriptaddress option in \documentclass). 
% \noaffiliation is required (may also be used with the \author command).
%\collaboration{}
%\noaffiliation

\date{\today}

\begin{abstract}
% insert abstract here
We show the stepped-pressure equilibria that are obtained from a generalization of Taylor relaxation known as multi-region, relaxed MHD (MRXMHD) are also generalizations of ideal MHD. We show this by proving that as the number of plasma regions becomes infinite, MRXMHD reduces to ideal MHD.  Numerical convergence studies demonstrating this limit are presented.
\end{abstract}

% insert suggested PACS numbers in braces on next line

\maketitle %\maketitle must follow title, authors, abstract and \pacs

% Body of paper goes here. Use proper sectioning commands. 
% References should be done using the \cite, \ref, and \label commands
\section{Introduction}

The construction of magnetohydrodynamic (MHD) equilibria in three-dimensional (3D) configurations is of fundamental importance for understanding toroidal magnetically confined plasmas.  The theory and numerical construction of 3D equilibria is complicated by the fact that toroidal magnetic fields without a continuous symmetry are generally a fractal mix of islands, chaotic field lines, and magnetic flux surfaces.  \citet{Hole:2007} have proposed a mathematically rigorous model for 3D MHD equilibria that embraces this structure by abandoning the assumption of continuously nested flux surfaces usually made when applying ideal MHD.  Instead a finite number of flux surfaces are assumed to exist in a partially-relaxed plasma system.  This model, termed a multiple relaxed region MHD (MRXMHD) model, is based on a generalization of the Taylor relaxation model \citep{Taylor:1974,Taylor:1986} in which the total energy (field plus plasma) is minimized subject to a finite number of magnetic flux, helicity and thermodynamic constraints.  The model leads to a stepped pressure profile, with the pressure jumps across the barrier interfaces counterbalanced by corresponding jumps in the magnitude of the magnetic field.

Although it might be expected that this MRXMHD model would reduce to ideal MHD in the limit of continuously nested flux surfaces, the discontinuous stepped-pressure profiles exhibited by this model make this unintuitive.  In this paper we prove that the MRXMHD model does reduce to ideal MHD in the limit of continuously nested flux surfaces and provide supporting numerical evidence using the Stepped Pressure Equilibrium Code (SPEC) \citep{Hudson:2012}.  This demonstrates that the model proposed by \citet{Hole:2007} reduces to usual results such as ideal MHD in the integrable limit where continuously nested flux surfaces exist.

In the next section we give a summary of the MRXMHD model and its solution for a finite number of plasma regions.  In Section~\ref{sec:ContinuouslyNestedSurfaceLimit} we prove the main result of the paper, that MRXMHD reduces to ideal MHD in the limit of continuously nested flux surfaces.  This is followed by supporting numerical evidence examining the convergence of SPEC to axisymmetric continuous pressure-profile solutions in Section~\ref{sec:NumericalConvergence}.  The paper is concluded in Section~\ref{sec:Conclusion}.

\section{The multiple-region relaxed MHD model}
\label{sec:MRXMHDModel}

\begin{figure}
  \includegraphics[width=8cm]{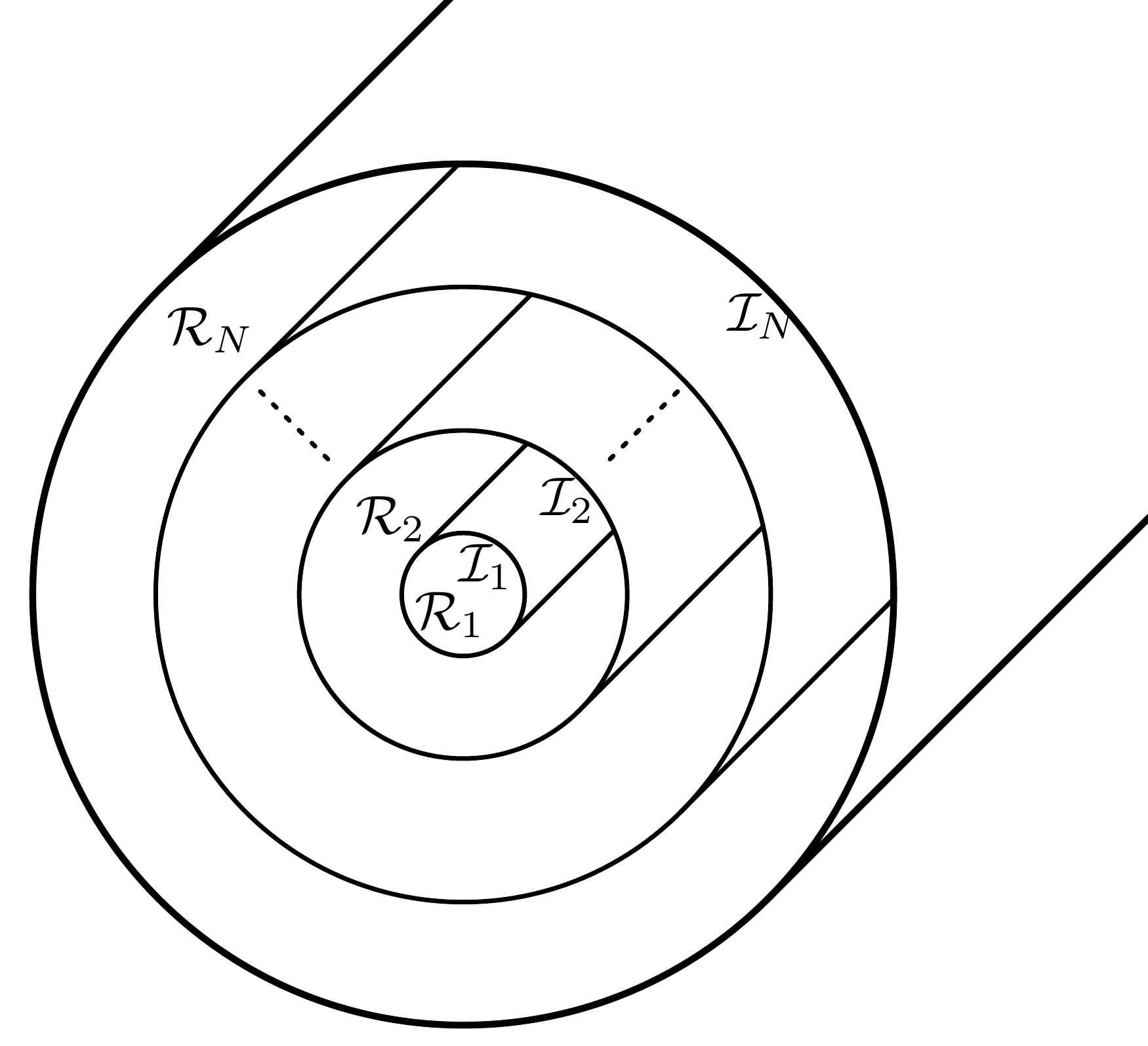}%
  \caption{Schematic of magnetic geometry showing ideal MHD barriers $\mathcal{I}_i$, and the relaxed plasma regions $\mathcal{R}_i$.\label{fig:NestedSurfaces}}%
\end{figure}

As introduced previously\citep{Hole:2006,Hole:2007,Hudson:2007,Dewar:2008} the MRXMHD model consists of $N$ nested plasma regions $\mathcal{R}_i$ separated by ideal MHD barriers $\mathcal{I}_i$ (see Fig.~\ref{fig:NestedSurfaces}).  Each plasma region is assumed to have undergone Taylor relaxation\citep{Taylor:1974} to a minimum energy state subject to conserved fluxes and magnetic helicity.  The energy functional for the MRXMHD model can be written as
\begin{align}
  W &= \sum_{i=1}^{N} U_i - \frac{1}{2}\sum_{i=1}^{N} \mu_i \left(H_i - H_i^{0}\right) - \sum_{i=1}^{N} \nu_i \left(S_i - S_i^{0}\right), \label{eq:FiniteInterfacesEnergyFunctional}
\end{align}
where there are $N$ nested plasma volumes, $\mu_i$ and $\nu_i$ are Lagrange multipliers, and
\begin{align}
  U_i &= \int_{\mathcal{R}_i} d\tau^3\, \left(\frac{p_i}{\gamma - 1} + \frac{1}{2}B_i^2 \right), \\
  S_i &= \int_{\mathcal{R}_i} d\tau^3\, p_i^{1/\gamma}, \label{eq:PlasmaEntropy} \\
  \begin{split}
  H_i &= \int_{\mathcal{R}_i} d\tau^3\, \mathbf{A} \cdot \mathbf{B} \\
  &\mathrel{\phantom{=}}- \Delta \psi_{p,i} \oint_{\mathcal{C}_{p,i}^{<}} d\mathbf{l} \cdot \mathbf{A} - \Delta \psi_{t,i} \oint_{\mathcal{C}_{t,i}^{>}} d\mathbf{l} \cdot \mathbf{A}. \label{eq:GaugeInvariantHelicity}
  \end{split}
\end{align}
In each plasma region $\mathcal{R}_i$ the term $U_i$ is the potential energy, $S_i$ the plasma entropy, and $H_i$ the gauge-invariant magnetic helicity\footnote{The form of the magnetic helicity in \eqref{eq:GaugeInvariantHelicity} differs from that in \citet{Hole:2007}, which is not fully gauge-invariant.}.  The plasma regions $\mathcal{R}_i$ are enclosed by flux surfaces, and are constrained to have helicity $H_i^0$, plasma entropy $S_i^0$, poloidal flux $\Delta\psi_{p,i}$ and toroidal flux $\Delta\psi_{t,i}$.  The $\mathcal{C}_{p,i}^{<}$ and $\mathcal{C}_{t,i}^{>}$ are circuits about the inner ($<$) and outer ($>$) boundaries of $\mathcal{R}_i$ in the poloidal and toroidal directions, respectively.

Local minimum energy solutions to \eqref{eq:FiniteInterfacesEnergyFunctional}--\eqref{eq:GaugeInvariantHelicity} are obtained by requiring the variation of $W$ to be zero.  With a fixed outer boundary $\mathcal{I}_N$, these solutions have the form\citep{Hole:2006,Hole:2007}
\begin{align}
  \mathcal{R}_i:& & \nabla \times \mathbf{B} &= \mu_i \mathbf{B}, & p_i &= \text{const}, \label{eq:FiniteInterfaceVolumeSolution}\\
  \mathcal{I}_i:& & \mathbf{n}\cdot\mathbf{B} &= 0, & \left[\left[p_i + \frac{1}{2} B^2 \right]\right] &= 0, \label{eq:FiniteInterfaceSurfaceSolution}
\end{align}
where \eqref{eq:FiniteInterfaceVolumeSolution} applies in each plasma region $\mathcal{R}_i$, \eqref{eq:FiniteInterfaceSurfaceSolution} applies on each ideal interface $\mathcal{I}_i$, $\mathbf{n}$ is a unit vector normal to the plasma interface $\mathcal{I}_i$ (see Figure~\ref{fig:NestedSurfaces}), and $[[x]] = x_{i+1} - x_{i}$ denotes the change in quantity $x$ across the interface $\mathcal{I}_i$.

\section{The continuously nested flux-surface limit}
\label{sec:ContinuouslyNestedSurfaceLimit}

In this section we show that the MRXMHD model reduces to ideal MHD as the number of plasma regions increases.  We begin by obtaining the limit of the MRXMHD energy functional \eqref{eq:FiniteInterfacesEnergyFunctional}--\eqref{eq:GaugeInvariantHelicity} for continuously nested flux surfaces.

We take the continuously nested flux surface limit of the MRXMHD energy functional \eqref{eq:FiniteInterfacesEnergyFunctional}--\eqref{eq:GaugeInvariantHelicity} by taking the limit as the number of plasma volumes $N\rightarrow \infty$ and the volume and enclosed fluxes of each plasma region approach zero.  In this limit the MRXMHD energy functional becomes
\begin{align}
  \begin{split}
  W =& \int d\tau^3\, \left(\frac{p}{\gamma - 1} + \frac{1}{2}B^2\right) \\
      &- \frac{1}{2}\int \mu(s) \left(dH - dH^0\right) - \int \nu(s) \left(dS - dS^0 \right), \label{eq:InitialInfiniteInterfaceMRXMHDFunctional}
  \end{split}
\end{align}
where $s$ is an arbitrary flux-surface label, $dH$ and $dS$ are the infinitesimal amounts of helicity and plasma entropy respectively between infinitesimally separated flux surfaces, and $dH^0$ and $dS^0$ are the corresponding constraints.  This energy functional is completed by expressions for the infinitesimal helicity $dH$ and plasma entropy $dS$.

The infinitesimal helicity $dH$ follows from \eqref{eq:GaugeInvariantHelicity},
\begin{align}
  dH &= d\tau^3\, \mathbf{A}\cdot\mathbf{B} - d\psi_p \oint_{\mathcal{C}_p(s)} d\mathbf{l}\cdot\mathbf{A} - d\psi_t \oint_{\mathcal{C}_t(s)} d\mathbf{l}\cdot\mathbf{A}, \label{eq:InfinitesimalHelicity}
\end{align}
where $\mathcal{C}_{t}(s)$ and $\mathcal{C}_{p}(s)$ are toroidal and poloidal circuits along flux surface $s$.  This may be further simplified by defining the enclosed flux functions
\begin{align}
  \psi_t(s) &= \oint_{\mathcal{C}_{p}(s)} d\mathbf{l}\cdot\mathbf{A} \label{eq:ToroidalFlux}, \\
  \psi_p(s) &= - \oint_{\mathcal{C}_{t}(s)} d\mathbf{l}\cdot\mathbf{A}, \label{eq:PoloidalFlux}
\end{align}
where $\psi_t(s)$ and $\psi_p(s)$ are the toroidal and poloidal fluxes enclosed by the flux surface $s$.

Using \eqref{eq:InfinitesimalHelicity}--\eqref{eq:PoloidalFlux} and the infinitesimal for $dS$ with \eqref{eq:InitialInfiniteInterfaceMRXMHDFunctional} gives the infinite-interface MRXMHD energy functional as
\begin{align}
  \begin{split}
    W =& \int d\tau^3 \, \left[\frac{p}{\gamma - 1} + \frac{1}{2}B^2 - \frac{1}{2} \mu(s) \mathbf{A}\cdot\mathbf{B} - \nu(s) p^{1/\gamma} \right] \\
    &+ \frac{1}{2} \int ds\, \mu(s)\left[ \psi_t(s) \frac{d\psi_p(s)}{ds} - \frac{d\psi_t(s)}{ds}\psi_p(s)\right] \\
    & + \int ds\, \left[\frac{1}{2}\mu(s) \frac{dH^{0}(s)}{ds} +  \nu(s) \frac{dS^0(s)}{ds} \right],
  \end{split} \label{eq:InfiniteInterfaceMRXMHDFunctional}
\end{align}
where $H^0(s)$ and $S^0(s)$ are the helicity and plasma entropy constraints.

The variation of this energy functional is subject to the constraints \eqref{eq:ToroidalFlux}--\eqref{eq:PoloidalFlux} on the poloidal and toroidal fluxes enclosed by each magnetic surface.  As discussed by \citet{Spies:2001}, these constraints lead to the following relationship between the variation of the vector potential $\delta\mathbf{A}$ and the variation of the interface positions $\delta\mathbf{x}$,
\begin{align}
  \mathbf{n}\times \delta \mathbf{A} &= - \left(\mathbf{n} \cdot \delta\mathbf{x}\right) \mathbf{B}, \label{eq:PotentialPositionConstraint}
\end{align}
where $\mathbf{n}$ is a unit vector normal to the flux surface.

In the next section we first reproduce the result of \citet{Taylor:1974} demonstrating that in the absence of pressure the time-independent solutions of \eqref{eq:InfiniteInterfaceMRXMHDFunctional} are nonlinear Beltrami fields. This result is then generalized to non-zero pressure in Section~\ref{sec:NonZeroPressure}.

\subsection{Zero pressure limit}

The zero-pressure limit of \eqref{eq:InfiniteInterfaceMRXMHDFunctional} may be taken by setting $p\rightarrow 0$, $\nu(s)\rightarrow 0$.  In this limit, we need to consider the variation of this functional with respect to the vector potential, the positions of the flux surfaces, and the Lagrange multiplier $\mu(s)$.

The variation $\delta\mu(s)$ is independent of $\delta\mathbf{A}$ and $\delta\mathbf{x}$, and may therefore be considered separately.  Requiring the variation of $W$ with respect to $\mu(s)$ be zero enforces the helicity constraint on each flux surface,
\begin{align}
  \left.\delta W\right|_\mu &= -\frac{1}{2} \int \delta\mu(s) \left(dH - dH^0\right) = 0,
\end{align}
or equivalently, $H(s) = H^0(s)$.

The remaining variation of $W$ with $p=0$ is
\begin{align}
  \begin{split}
    \delta W =& \int d\tau^3 \, \left[\mathbf{B}\cdot\delta\mathbf{B} - \frac{1}{2} \mu(s) \left(\delta\mathbf{A}\cdot\mathbf{B} + \mathbf{A}\cdot\delta\mathbf{B}\right)\right] \\
    &- \int d\tau^3 \frac{1}{2} \mathbf{A}\cdot\mathbf{B} \frac{d\mu(s)}{ds} \delta s(\mathbf{x}),
  \end{split} \label{eq:InfiniteInterfaceMRXMHDFunctionalZeroPressure}
\end{align}
where $s(\mathbf{x})$ is the flux surface label as a function of position.  The variation of the terms on the second line of \eqref{eq:InfiniteInterfaceMRXMHDFunctional} with fixed $\mu(s)$ is zero as $\psi_t(s)$ and $\psi_p(s)$ are given functions of the flux surface label $s$.

The variation of $s(\mathbf{x})$ can be obtained by defining $\tilde{s}(\mathbf{x}) \equiv s(\mathbf{x}) + \delta s(\mathbf{x})$ to be the flux surface label after interface perturbation, and using the requirement that the perturbation doesn't change the label of a flux surface:
\begin{align}
  \tilde{s}(\mathbf{x} + \delta \mathbf{x}) &= s(\mathbf{x}), \\
  s(\mathbf{x}) + \delta s(\mathbf{x}) + \delta \mathbf{x} \cdot \nabla s(\mathbf{x}) &= s(\mathbf{x}), \\
  \delta s(\mathbf{x}) &= - \delta\mathbf{x} \cdot \nabla s(\mathbf{x}). \label{eq:SPerturbation}
\end{align}

Using \eqref{eq:SPerturbation}, the energy functional \eqref{eq:InfiniteInterfaceMRXMHDFunctional} may be written as
\begin{align}
  \begin{split}
    \delta W =& \int d\tau^3 \, \left[\mathbf{B}\cdot\nabla \times \delta\mathbf{A} - \frac{1}{2} \mu(s) \left(\delta\mathbf{A}\cdot\mathbf{B} + \mathbf{A}\cdot\nabla \times\delta\mathbf{A}\right)\right] \\
    &+ \frac{1}{2}\int d\tau^3 \mathbf{A}\cdot\mathbf{B} \left(\delta\mathbf{x}\cdot\nabla \mu\right),
  \end{split} \label{eq:InfiniteInterfaceMRXMHDFunctionalZeroPressure2}
\end{align}
where the perturbation of the magnetic field $\delta\mathbf{B}$ has been written in terms of the perturbation of the vector potential using $\delta\mathbf{B} = \nabla \times \delta\mathbf{A}$.

This expression may be simplified using the relation
\begin{align}
  \int d\tau^3\, \mathbf{Q} \cdot \nabla \times \delta\mathbf{A} &= \int d\tau^3\, \delta\mathbf{A} \cdot \nabla \times \mathbf{Q}, \label{eq:VectorIdentity}
\end{align}
where $\mathbf{Q}$ is an arbitrary single-valued vector field, and Eq.~\eqref{eq:PotentialPositionConstraint} and the assumption that the outermost interface remain fixed (i.e.\ $\mathbf{n}\cdot\delta\mathbf{x}=0$ on the boundary) have been used.

The relation                   \eqref{eq:VectorIdentity} may now be used with $\mathbf{Q}\rightarrow\mathbf{B}$ and $\mathbf{Q}\rightarrow\mu\mathbf{A}$ to simplify \eqref{eq:InfiniteInterfaceMRXMHDFunctionalZeroPressure2},
\begin{align}
  \begin{split}
    \delta W =& \int d\tau^3 \, \delta\mathbf{A} \cdot \left\{\nabla \times\mathbf{B} - \frac{1}{2} \left[\mu \mathbf{B} + \nabla \times (\mu \mathbf{A}) \right] \right\} \\
    &+ \frac{1}{2}\int d\tau^3 \mathbf{A}\cdot\mathbf{B} \left(\delta\mathbf{x}\cdot\nabla \mu\right),
  \end{split} \\
  \begin{split}
    \delta W =& \int d\tau^3 \, \delta\mathbf{A} \cdot \left[\nabla\times\mathbf{B} - \mu(s) \mathbf{B}\right] \\
    &+ \frac{1}{2}\int d\tau^3 \left[\mathbf{A}\cdot\mathbf{B} \left(\delta\mathbf{x}\cdot\nabla \mu\right) - \left(\nabla \mu \times \mathbf{A}\right) \cdot\delta\mathbf{A} \right].
  \end{split} \label{eq:InfiniteInterfaceMRXMHDFunctionalZeroPressure3}
\end{align}
The last line of \eqref{eq:InfiniteInterfaceMRXMHDFunctionalZeroPressure3} is zero:
\begin{align}
  \int d\tau^3\, \mathbf{A} \cdot\mathbf{B} (\delta\mathbf{x} \cdot \nabla \mu) &= -\int d\tau^3\, \mathbf{A} \cdot \left(\nabla \mu \times \delta\mathbf{A}\right), \\
  &= \int d\tau^3\, \left(\nabla \mu \times \mathbf{A} \right) \cdot \delta\mathbf{A},
\end{align}
where \eqref{eq:PotentialPositionConstraint} has been used, noting that $\nabla \mu(s) \parallel \mathbf{n}$.

The variation $\delta W$ has now been shown to be
\begin{align}
  \delta W =& \int d\tau^3 \, \delta\mathbf{A} \cdot \left[\nabla\times\mathbf{B} - \mu(s) \mathbf{B}\right]. \label{eq:InfiniteInterfaceMRXMHDFunctionalZeroPressure4}
\end{align}

It is tempting to conclude from \eqref{eq:InfiniteInterfaceMRXMHDFunctionalZeroPressure4} that $\nabla \times \mathbf{B} = \mu(s) \mathbf{B}$, however this is not true in general.  The flux conservation condition \eqref{eq:PotentialPositionConstraint} requires that $\delta \mathbf{A} \cdot\mathbf{B} = 0$, hence $\delta \mathbf{A}$ is not a completely free variation.  Requiring that the energy variation $\delta W$ in \eqref{eq:InfiniteInterfaceMRXMHDFunctionalZeroPressure4} be zero for all possible variations only shows that the coefficients of \emph{independent} variations are zero.

The potential variation $\delta\mathbf{A}$ can be written in terms of independent variations using \eqref{eq:PotentialPositionConstraint},
\begin{align}
  \delta \mathbf{A} &= \delta \mathbf{x} \times \mathbf{B} + \mathbf{n}\, \delta A_\perp, \label{eq:dAdxRelation}
\end{align}
where $\delta A_\perp$ is the remaining free variation of $\mathbf{A}$, which is perpendicular to the flux surfaces.  $\delta A_\perp$ is independent of $\delta\mathbf{x}$.

Using \eqref{eq:dAdxRelation}, the energy functional variation \eqref{eq:InfiniteInterfaceMRXMHDFunctionalZeroPressure4} may be written as
\begin{align}
  \delta W =& \int d\tau^3 \, \left[ \delta \mathbf{x} \cdot \left( - \mathbf{J} \times \mathbf{B} \right) + \delta A_\perp \mathbf{n} \cdot \mathbf{J} \right]. \label{eq:InfiniteInterfaceMRXMHDFunctionalZeroPressureFinal}
\end{align}
As $\delta \mathbf{x}$ and $\delta A_\perp$ are independent, the time-independent solutions satisfy
\begin{align}
  \mathbf{J} \times \mathbf{B} &= 0, \\
  \mathbf{n} \cdot \mathbf{J} &= 0.
\end{align}
These two conditions imply that the current is parallel to the magnetic field
\begin{align}
  \nabla \times \mathbf{B} &= \lambda(\mathbf{x}) \mathbf{B},
\end{align}
for some $\lambda(\mathbf{x})$.  As the fields and currents are time-independent $\nabla\cdot\mathbf{J} = 0$ implies that $\mathbf{B}\cdot \nabla\lambda = 0$, hence $\lambda$ is constant on a field line.  

Time-independent solutions of the infinite interface limit of the MRXMHD model without pressure are therefore nonlinear Beltrami fields
\begin{align}
  \nabla \times \mathbf{B} &= \lambda(\alpha) \mathbf{B}, \label{eq:NonlinearBeltramiSolution}
\end{align}
where $\alpha$ labels the field line.  This is the result of \citet{Taylor:1974}.  

One might have expected $\mu(s)$ to replace $\lambda(\alpha)$ in \eqref{eq:NonlinearBeltramiSolution} because for a finite number of interfaces the plasma in each volume satisfies $\nabla\times\mathbf{B} = \mu_i \mathbf{B}$ [see \eqref{eq:FiniteInterfaceVolumeSolution}].  However, there are also surface currents on the interfaces between the plasma volumes.  In the limit of an infinite number of continuously nested surfaces, the plasma volume current will have contributions both from the volume and surface currents of the finite-$N$ case.  Only if the surface currents in the finite-$N$ case are zero should we expect $\lambda(\alpha)$ to be equal to $\mu(s)$.  For example, the surface currents will be zero if the $\mu_i$ are all equal, and in this case the solution is
\begin{align}
  \nabla \times \mathbf{B} &= \mu \mathbf{B},
\end{align}
with $\mu$ a constant.  In this case $\lambda(\alpha)$ is equal to $\mu(s)$.

% How would one get the surface currents to be zero?  This requires that the magnetic field is continuous across the interfaces.  Certainly that is possible if $\mu(s)$ = const, but is there a more general solution?  This requires that the rotational transform be continuous.

In the next section we consider the effect of pressure on the time-independent solutions of the infinite interface MRXMHD model.

\subsection{Non-zero pressure}
\label{sec:NonZeroPressure}

For non-zero pressure, the additional terms to the variation \eqref{eq:InfiniteInterfaceMRXMHDFunctionalZeroPressureFinal} that must be considered are the variations of the $\nu$ and $p$ terms in \eqref{eq:InfiniteInterfaceMRXMHDFunctional}.

The variation with respect to $\nu(s)$ enforces the plasma entropy constraint
\begin{align}
  \left.\delta W \right|_\nu &= - \int \delta\nu(s)\, \left(dS - dS^0 \right) = 0,
\end{align}
or $S(s) = S^0(s)$.

The variation with respect to pressure is
\begin{align}
  \left.\delta W\right|_p &= \int \delta p \left[ \frac{1}{\gamma - 1} - \nu(s) \frac{1}{\gamma} p^{1/\gamma - 1} \right].
\end{align}
As the variation $\delta p$ is independent of $\delta\mathbf{x}$ and $\delta\mathbf{A}$, time-independent solutions satisfy
\begin{align}
  \nu(s) p^{1/\gamma - 1} &= \frac{\gamma}{\gamma - 1}, \label{eq:PressureOnFluxSurface}
\end{align}
which implies that $p$ is constant on a flux surface.

The remaining additional term to the energy variation in the zero-pressure case is the variation of $\nu(s)$ as the interface positions are varied.  This term is
\begin{align}
  \int d\tau^3 p^{1/\gamma} (\delta \mathbf{x} \cdot \nabla \nu). \label{eq:NonZeroPressureAdditionalTerm}
\end{align}
The gradient of $\nu$ can be written in terms of the pressure $p$ using \eqref{eq:PressureOnFluxSurface},
\begin{align}
  p^{1/\gamma} \nabla \nu &= \nabla p. \label{eq:NuPressureGradientRelation}
\end{align}

The variation of the MRXMHD energy functional including pressure is \eqref{eq:InfiniteInterfaceMRXMHDFunctionalZeroPressureFinal} with the additional term \eqref{eq:NonZeroPressureAdditionalTerm}
\begin{align}
  \delta W =& \int d\tau^3 \, \left[ \delta \mathbf{x} \cdot \left(\nabla p - \mathbf{J} \times \mathbf{B} \right) + \delta A_\perp \mathbf{n} \cdot \mathbf{J} \right], \label{eq:InfiniteInterfaceMRXMHDFunctionalFinal}
\end{align}
where \eqref{eq:NuPressureGradientRelation} has been used. As the variations $\delta \mathbf{x}$ and $\delta A_\perp$ are independent, the time-independent solutions of the infinite interface MRXMHD functional satisfy
\begin{align}
  \mathbf{J} \times \mathbf{B} &= \nabla p, \\
  \mathbf{n} \cdot \mathbf{J} &= 0,
\end{align}
which are the equations for ideal MHD.  In the limit of continuously nested flux surfaces, MRXMHD is equivalent to ideal MHD.  In particular, in the axisymmetric $N\rightarrow\infty$ limit MRXMHD reduces to the Grad-Shafranov equation.

In the next section we use SPEC\citep{Hudson:2012} to illustrate the convergence of the MRXMHD model to axisymmetric ideal MHD with continuous pressure profiles.

\section{Numerical illustration}
\label{sec:NumericalConvergence}

The Stepped Pressure Equilibrium Code\citep{Hudson:2012} solves the MRXMHD model \eqref{eq:FiniteInterfacesEnergyFunctional}--\eqref{eq:GaugeInvariantHelicity} for an arbitrary (finite) number of plasma regions.  We use this code to illustrate the results of the previous section by showing the numerical convergence of SPEC to an axisymmetric continuous pressure-profile ideal MHD solution as computed by the VMEC code\citep{Hirshman:1998}.

The equilibrium is defined by a given, fixed outer boundary, the pressure and rotational-transform profiles as a function of the normalized toroidal flux, $s = \psi/\psi_\text{encl}$, where $\psi_\text{encl}$ is the total enclosed toroidal flux.  For this comparison we choose $\psi_\text{encl} = 2\pi$ in units where $\mu_0=1$.

For the numerical convergence study we choose the fixed outer boundary to be an axisymmetric torus with circular cross-section:
\begin{eqnarray}
R = 1.0 + 0.3 \cos(\theta), \; Z =       0.3 \sin(\theta).
\end{eqnarray}
We define the equilibrium by choosing the pressure and rotational transform flux functions.  The continuous pressure profile is selected to be
\begin{eqnarray}
p(s) = p_0( 1 - 2 s + s^2), \label{eq:BenchmarkPressureProfile}
\end{eqnarray}
where $p_0$ is to be adjusted; e.g. $p_0=0$ for zero-beta.  The continuous transform profile is selected to be
\begin{eqnarray}
\iota(s) = \iota_a + (\iota_e-\iota_a) s,
\end{eqnarray}
where $\iota_a = (8+\gamma 9)/(9+\gamma 10)$ and $\iota_e = (1+\gamma 1)/(9+\gamma 10)$, and $\gamma\equiv ( 1 + \sqrt 5 ) / 2$ is the golden mean.  This transform profile is selected to ensure that the rotational transform on the ideal interfaces in the MRXMHD model are noble irrationals.  This ensures stability of the ideal interfaces\citep{McGann:2010}.

\begin{figure}
 \includegraphics[width=8cm]{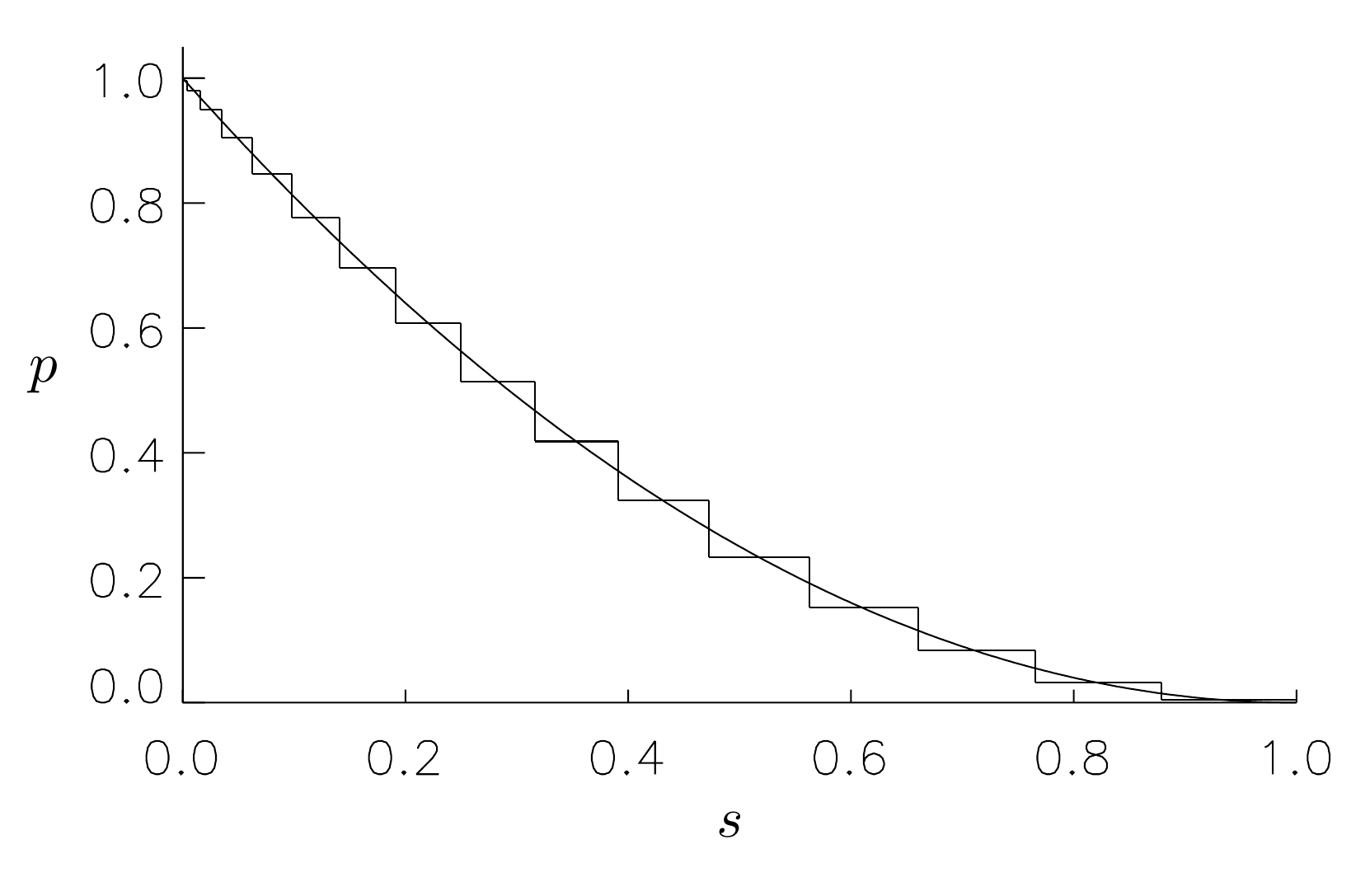}%
 \caption{Pressure profile used for demonstrating the $N\rightarrow\infty$ limit of MRXMHD.  The continuous curve is the pressure profile in \eqref{eq:BenchmarkPressureProfile} which is used with VMEC, and the stepped profile is the $N=16$ approximation used with SPEC.\label{fig:discretepressureprofile}}%
 \end{figure}

As input to SPEC, these profiles are discretized as follows.
For convergence studies as the number of plasma regions $N\rightarrow \infty$, is it convenient to have the SPEC interfaces equally spaced in $\sqrt s$.
So, for $i=1, \dots, N$, we define $s_i \equiv \sqrt{i/N}$ and the interface transforms as $\iota_i=\iota(s_i)$.
The pressure in each volume is constructed so that 
\begin{eqnarray}
p_i \int_{s_{i-1}}^{s_{i}} \!\!\! ds = \int_{s_{i-1}}^{s_{i}} \!\!\! p(s) \, ds.
\end{eqnarray}
A discrete pressure profile, with $N=16$ is shown in Fig.~\ref{fig:discretepressureprofile}.

\begin{figure}
 \includegraphics{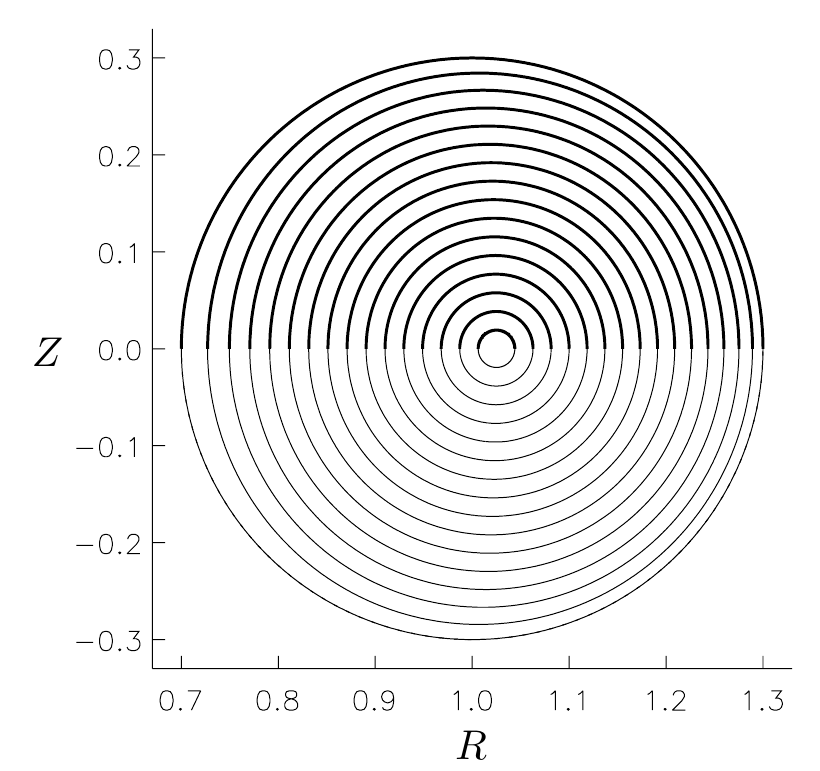}%
 \caption{Comparison of magnetic surfaces: zero-pressure (\mbox{$p_0=0$}); the SPEC interfaces are shown in thin lines (upper and lower half) and the VMEC surfaces are shown in thick lines (upper half only).\label{fig:zerobetapoincare}}%
 \end{figure}

\begin{figure}
 \includegraphics{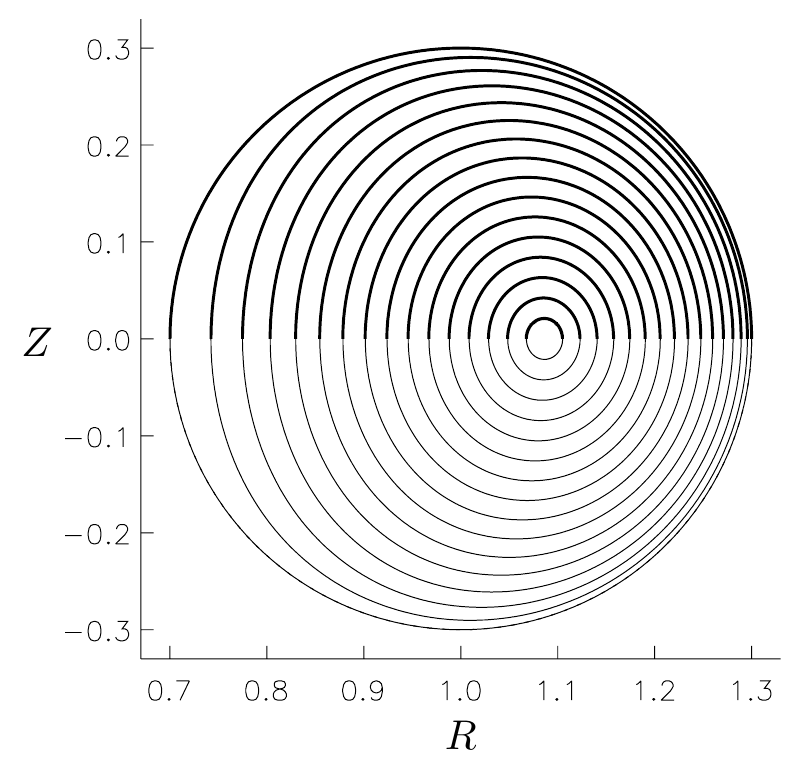}
 \caption{Comparison of magnetic surfaces: high-pressure (\mbox{$p_0=16$}); the SPEC interfaces are shown in thin lines (upper and lower half) and the VMEC surfaces are shown in thick lines (upper half only).\label{fig:highbetapoincare}}%
 \end{figure}

A comparison of the SPEC interfaces, for an $N=16$, zero-$\beta$ case (i.e. $p_0=0$), is shown in Fig.~\ref{fig:zerobetapoincare}, and for a high-$\beta$ case in Fig.~\ref{fig:highbetapoincare}.
For the high-pressure case, $p_0$ was increased to give a Shafranov shift about one third the minor radius.

To quantify the difference between the SPEC and VMEC solutions, we define a measure of the difference in geometry of a given magnetic surface as
\begin{align}
  \Delta &\equiv \int_{0}^{2\pi} d\theta \left| \mathbf{x}_\text{VMEC}(\theta)-\mathbf{x}_\text{SPEC}(\theta)\right|,
\end{align}
where $\mathbf{x}(\theta)$ is the intersection of the surface with the $\phi=0$ plane, and $\theta$ is the polar angle.

\begin{figure}
 \includegraphics{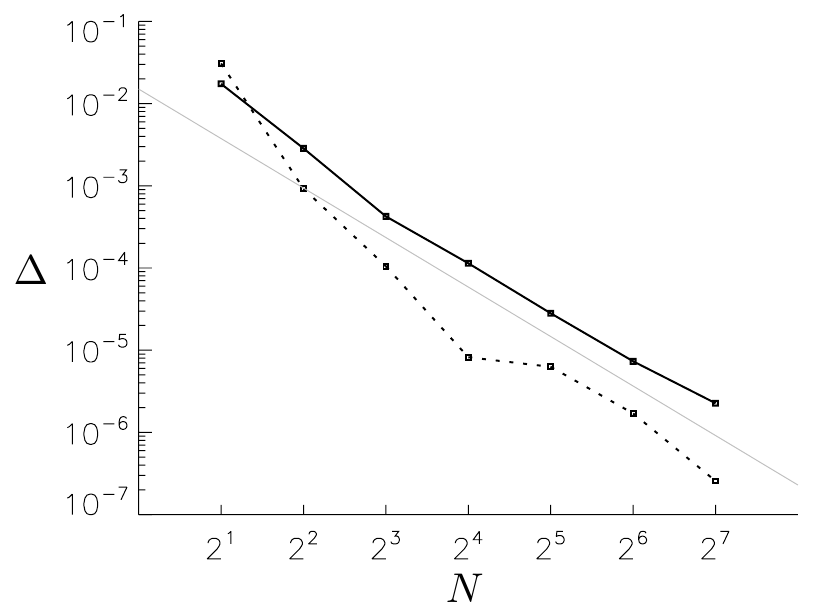}
 \caption{Convergence: The error ($\Delta$) between the continuous pressure (VMEC) and stepped pressure (SPEC) solutions are shown as a function of the number of plasma regions $N$ for the $s=1/4$ SPEC interface.  The dotted line shows the zero-beta case ($p_0=0$), and the solid line shows the high-beta case ($p_0=16$). The grey line has a slope $-2$, the expected rate of convergence.\label{fig:Nvlimit}}%
 \end{figure}

Figure~\ref{fig:Nvlimit} shows $\Delta$ computed between the representative $s=1/4$ SPEC interface and the corresponding VMEC interface as the number of interfaces $N$ is increased up to the maximum afforded by computational limitations and expedience of $N=128$.  In particular, the convergence of the error is second order, $\Delta \sim N^{-2}$, as shown in Fig.~\ref{fig:Nvlimit}.

\section{Conclusion}
\label{sec:Conclusion}

We have demonstrated that the Multiple-Region Relaxed MHD model reduces to ideal MHD in the limit of an infinite number of plasma regions.  In this limit, the magnetic geometry is characterized by continuously nested flux surfaces. The appeal of MRXMHD is that for a finite number of plasma regions, only a finite number of flux surfaces are assumed to exist.  The rest of the plasma may be characterized by smoothly nested flux surfaces, islands, chaotic fields, or some combination of these.  In particular, the work of \citet{Hudson:2012} demonstrates the application of SPEC to a DIIID equilibrium with a fully 3D boundary in which magnetic islands form.  In future work we will apply MRXMHD and SPEC to the RFX Quasi-Single Helicity state \citep{Lorenzini:2009} in which two magnetic axes have been shown to form.

\section*{Acknowledgements}

The authors gratefully acknowledge support of the U.S. Department of Energy and the Australian Research Council, through Grants DP0452728, FT0991899, and DP110102881.

% If in two-column mode, this environment will change to single-column format so that long equations can be displayed. 
% Use only when necessary.
%\begin{widetext}
%$$\mbox{put long equation here}$$
%\end{widetext}

% Figures should be put into the text as floats. 
% Use the graphics or graphicx packages (distributed with LaTeX2e).
% See the LaTeX Graphics Companion by Michel Goosens, Sebastian Rahtz, and Frank Mittelbach for examples. 
%
% Here is an example of the general form of a figure:
% Fill in the caption in the braces of the \caption{} command. 
% Put the label that you will use with \ref{} command in the braces of the \label{} command.
%
% \begin{figure}
% \includegraphics{}%
% \caption{\label{}}%
% \end{figure}

% Tables may be be put in the text as floats.
% Here is an example of the general form of a table:
% Fill in the caption in the braces of the \caption{} command. Put the label
% that you will use with \ref{} command in the braces of the \label{} command.
% Insert the column specifiers (l, r, c, d, etc.) in the empty braces of the
% \begin{tabular}{} command.
%
% \begin{table}
% \caption{\label{} }
% \begin{tabular}{}
% \end{tabular}
% \end{table}

% If you have acknowledgments, this puts in the proper section head.
%\begin{acknowledgments}
% Put your acknowledgments here.
%\end{acknowledgments}

% Create the reference section using BibTeX:
\bibliography{MRXMHD_Limit_Paper_Final.bib}

\end{document}